\documentclass[showpacs,preprintnumbers,amsmath,amssymb,twocolumn]{revtex4}

\usepackage{graphicx}
\usepackage{dcolumn}
\usepackage{bm}

\begin{document}

\title{Magnetic Stability Analysis for Abelian  and Non-Abelian Superconductors}
\author{\normalsize{Lianyi He\footnote{Email address: hely04@mails.tsinghua.edu.cn}$^1$,
                    Meng Jin\footnote{Email address: jin-m@mail.tsinghua.edu.cn}$^{1,2}$
                and Pengfei Zhuang\footnote{Email address: zhuangpf@mail.tsinghua.edu.cn}$^1$}}
\affiliation{1 Physics Department, Tsinghua University, Beijing 100084, China\\
             2 Institute of Particle Physics, Central China Normal University,
             Wuhan 430070, China}

\begin{abstract}
We investigate the origin of Abelian and non-Abelian type magnetic
instabilities induced by Fermi surface mismatch between the two
pairing fermions in a non-relativistic model. The Abelian type
instability occurs only in gapless state and the Meissner mass
squared becomes divergent at the gapless-gapped transition point,
while the non-Abelian type instability happens in both gapless and
gapped states and the divergence vanishes. The non-Abelian type
instability can be cured in strong coupling region.
\end{abstract}

\maketitle

\section {Introduction}
\label{s1}
The fermion Cooper pairing between different species with
mismatched Fermi surfaces, which was discussed in electronic
superconductivity many year ago\cite{sarma,LO,FF,takada}, promoted
new interest in the study of color superconducting quark matter in
compact stars\cite{huang,shovkovy,alford}, asymmetric nuclear
matter with proton-neutron pairing\cite{sedrakian,sedrakian2}, and
fermionic atom gas with population
imbalance\cite{petit,sheehy,exp1,exp2}.

For color superconducting quark matter with beta equilibrium and
charge neutrality, the Sarma\cite{sarma} or breached
pairing(BP)\cite{liu} or gapless\cite{shovkovy} state possesses
paramagnetic response to the external color magnetic fields, i.e.,
the Meissner masses squared of some gluons are
negative\cite{huang1,casalbuoni,alford1,fukushima}. Therefore, the
homogeneous and isotropic gapless state is dynamically unstable in
weak coupling BCS region, and some spatially inhomogeneous and
anisotropic states are energetically
favored\cite{giannakis,huang2,hong,gorbar,he1}.

Since the 8th gluon corresponds to the diagonal generator $T_8$ in
color space, and in an Abelian system the Meissner mass squared is
proportional to the superfluid density, the chromomagnetic
instability of the 8th gluon in the gapless two flavor color
superconductivity and the negative superfluid density in the BP
superfluidity are of the same type
instabilities\cite{wu,he2,gubankova}. Generally, the Abelian type
magnetic instability is related to the Fermi surface
topology\cite{gubankova}. In weak coupling region, one branch of
fermionic quasiparticle dispersions has two gapless nodes and
there are two sharp effective Fermi surfaces in momentum space. In
this case, the Meissner mass squared or the superfluid density is
negatively divergent at the gapless-gapped transition point where
the chemical potential mismatch between the two pairing fermions
is equal to the pairing gap. At strong enough coupling, the branch
of the gapless fermionic excitation has only one gapless node, and
therefore there is only one sharp effective Fermi surface in
momentum space. This type of gapless state is free from both Sarma
and magnetic instabilities\cite{pao,kitazawa,gubankova} and is
energetically more favored than the mixed phase\cite{caldas} and
LOFF phase\cite{LO,FF}.

There exists another type of magnetic instability for the 4-7th
gluons\cite{huang1} in the two flavor color superconductivity. We
call it non-Abelian magnetic instability. Very different from the
Abelian instability for the 8th gluon, the Meissner mass squared
for the 4-7th gluons is negative in both the gapless and gapped
states and is not divergent at the gapless-gapped transition
point. It is recently argued that the LOFF state suffers also
negative Meissner masses squared for the 4-7th gluons and the
gluonic phase will be energetically favored in a wide range of
coupling\cite{gorbar,kiriyama}.

In this paper, we will show that the non-Abelian instability is
related only to the non-Abelian gauge symmetry and is independent
of the details of the attractive interaction between the two
pairing fermions. Any system which possesses a SU(3) gauge
symmetry may suffer this type of magnetic instability. By
regarding the gauge field as the pseudo Nambu-Goldstone
current\cite{huang2}, the non-Abelian instability analysis can be
applied to the systems of condensed matter or fermionic atom gas
where there is no real non-Abelian gauge field. The paper is
organized as follows. We present the general treatment for a
non-relativistic fermion system in Section \ref{s2}, then
investigate the magnetic instabilities in Abelian and non-Abelian
cases in Sections \ref{s3} and \ref{s4}, respectively, and
summarize in Section \ref{s5}. We use the natural unit of
$c=\hbar=k_B=1$ through the paper.

\section {General Frame}
\label{s2}
For a non-relativistic fermion system, the partition function of
the system can be generally expressed as
\begin{equation}
Z=\int[d\psi^\dagger][d\psi]e^{\int_x\left[\psi^\dagger\left(-\partial_\tau+\frac{\nabla^2}{2m}+\mu\right)\psi+{\cal
L}_{int}\right]}
\end{equation}
in the imaginary time formalism of finite temperature field theory
with coordinates $x_\mu=\left(\tau=it,\ {\bf x}\right)$, where
$\int_x=\int d^3{\bf x}\int_0^\beta d\tau$ is the space-time
integration with $\beta=1/T$ related to the temperature $T$ of the
system, $\psi, m$ and $\mu$ are, respectively, the fermion field,
mass and chemical potential, and the interaction sector ${\cal
L}_{int}$ provides an attractive coupling between the pairing
fermions.

Since a fermion can possess possible inner degrees of freedom, the
quantities $\psi, m$ and $\mu$ are normally matrices in inner
space. To simplify the calculation, we neglect in this paper mass
difference and consider only chemical potential difference in the
inner space. Generally, due to the introduction of inner degrees
of freedom, the system may have some gauge symmetry, such as a
SU(3) gauge symmetry which can, for instance, be realized in a
$^{40}$K atom gas with three hyperfine states\cite{honerkamp}.
Suppose the system under consideration has a gauge symmetry with
$N$ generators $T_a(a=1,2,...,N)$, we can introduce a model gauge
field $A_{\mu}^a(a=1,2,...,N)$ corresponding to the gauge
symmetry, and the partition function including the gauge field can
then be written as
\begin{equation}
Z=\int[d\psi^\dagger][d\psi][dA]e^{S_A}e^{\int_x\left[\psi^\dagger\left(-D_\tau+\frac{{\bf
D}^2}{2m}+\mu\right)\psi+{\cal L}_{int}\right]},
\end{equation}
where the gauge field enters the theory via the standard gauge
coupling
\begin{equation}
\label{gauge}
D_\mu=\partial_\mu-igT_aA_\mu^a
\end{equation}
with the gauge coupling constant $g$, and $S_A$ is the action for
the gauge field sector which is not shown explicitly here.

Due to the attractive interaction between the fermions, the BCS
pairing occurs and breaks the gauge symmetry. The order parameter
field $\phi$ which describes the spontaneous symmetry breaking is
generally a matrix in the inner space. For $\Delta\equiv
\left<\phi\right>\neq 0$, the symmetry group of the Lagrangian is
spontaneously broken down to some subgroups, and some components
of the gauge field corresponding to the broken generators will
obtain the so-called Meissner mass via Higgs mechanism.

What we are interested in in this paper is how the Fermi surface
mismatch between the pairing fermions affects the behavior of the
Meissner mass, especially, whether it is real and physical. When
some Meissner masses squared become negative, the system suffers
magnetic instability, and the state with gauge field condensation
$\langle {\bf A}_a\rangle\neq 0$ which breaks the rotational
symmetry is energetically more favored than the homogeneous and
isotropic state.

The model gauge field $A_\mu^a$ we introduced above is in fact not
necessary for the magnetic instability analysis. If their is no
such a gauge field in the theory, for instance in a fermionic atom
gas, it can be replaced by the space-time derivative of the phase
fluctuation or the Nambu-Goldstone current\cite{huang2}
corresponding to the broken generators. Generally, we can define a
local phase transformation by
\begin{equation}
{\cal V}(x)=e^{i\sum_{a=1}^N\varphi_a(x)T_a},
\end{equation}
where $\varphi_a(x)$ can be regarded as the phase field related to
the generator $T_a$. Applying the above transformation to the
fermion field,
\begin{equation}
\psi(x) = {\cal V}(x)\chi(x),\ \ \ \psi^\dagger(x) =
\chi^\dagger(x){\cal V}^\dagger(x),
\end{equation}
we derive the partition function of the system in terms of the
fields $\chi$ and $\varphi$,
\begin{equation}
Z=\int[d\chi^\dagger][d\chi][d\varphi]e^{\int_x\left[\chi^\dagger\left(-{\cal
D}_\tau+\frac{{\bf {\cal D}}^2}{2m}+\mu\right)\chi+{\cal
L}_{int}\right]},
\end{equation}
where the derivative ${\cal D}_\mu$ is defined as
\begin{equation}
{\cal D}_\mu=\partial_\mu-iV_\mu(x)
\end{equation}
with the Nambu-Goldstone current
\begin{equation}
V_\mu(x)={\cal V}^\dagger(x)(i\partial_\mu){\cal V}(x).
\end{equation}
To the linear order in the phase field $\varphi$, the
Nambu-Goldstone current takes the form
\begin{equation}
V_\mu(x)=\sum_{a=1}^N\partial_\mu\varphi_a(x)T_a.
\end{equation}
From the comparison with the standard gauge coupling
(\ref{gauge}), the model gauge field $A_\mu^a$ and the nonzero
gauge field condensation $\langle{\bf A}_a\rangle\neq 0$ in
symmetry breaking state can be replaced, respectively, by the
Nambu-Goldstone current $\partial_\mu\varphi_a$ and its ensemble
average\cite{huang2},
\begin{equation}
\langle\nabla\varphi_a\rangle\neq 0.
\end{equation}
Therefore, even for those systems without real gauge field, the
calculation of the Meissner mass for the model gauge field is
still meaningful.

The gauge field condensation or spontaneously generated
Nambu-Goldstone current can be absorbed into a LOFF state where
the order parameter takes the form
\begin{equation}
\left<\phi(x)\right>=e^{-2iT_a{\bf q}_a\cdot{\bf x}}\Delta,
\end{equation}
if we identify the wave vector ${\bf q}_a=g\langle{\bf
A}_a\rangle=\langle\nabla\varphi_a\rangle$.

The thermodynamic potential $\Omega$ of the system is a function
of the order parameter $\Delta$ and pair momentum ${\bf q}_a$, and
the physical values of $\Delta$ and ${\bf q}_a$ at fixed
temperature and chemical potentials in the ground state correspond
to the minimum of the thermodynamic potential,
\begin{equation}
{\partial \Omega\over \partial \Delta}=0,\ \ \
{\partial^2\Omega\over\partial\Delta^2}>0,\ \ \ {\partial
\Omega\over \partial {\bf q}_a}=0,\ \ \
{\partial^2\Omega\over\partial {\bf q}_a^2}>0.
\end{equation}
Since $\Omega(\Delta,{\bf q}_a)$ in the LOFF state can be expanded
in powers of ${\bf q}_a$ at the vicinity of ${\bf q}_a=0$,
\begin{equation}
\Omega(\Delta,{\bf q}_a)=\Omega(\Delta,0)+
\frac{1}{2g^2}M^2_{ab}{\bf q}_a\cdot{\bf q}_b+...,
\end{equation}
the coefficients $M^2_{ab}$ are just the Meissner masses squared
matrix for the gauge field. If some of its elements are negative,
the LOFF state is energetically favored. For an Abelian system,
$T_a=1$, such a LOFF state is just the classical single-wave LOFF
state\cite{FF,takada}, and for a non-Abelian system, it can be
regarded as an extended non-Abelian LOFF
state\cite{fukus,hashimoto}.

\section {Abelian Gauge Symmetry}
\label{s3}
For an Abelian system with two species of fermions
($\psi=\left(\psi_1, \psi_2\right)^T$), a U(1) gauge field $A_\mu$
can be introduced into the system via the covariant derivative
\begin{equation}
D_\mu=\partial_\mu-ieA_\mu
\end{equation}
with gauge coupling $e$, and the attractive interaction between
the two species can be modeled by a point interaction with
coupling constant $G$,
\begin{equation}
{\cal L}_{int} = G\left(\psi^\dagger
i\varepsilon\psi^*\right)\left(\psi^Ti\varepsilon\psi\right),
\end{equation}
where we have assumed that the pairing can occur only between
different species of fermions, indicated by the antisymmetric
tensor
$(\varepsilon)_{\alpha\beta}\equiv\varepsilon_{\alpha\beta}$ in
the inner space. The order parameter describing spontaneous U(1)
symmetry breaking is defined as the expectation value of the pair
field,
\begin{equation}
\Delta=-2G\langle\psi^Ti\varepsilon\psi\rangle.
\end{equation}
In a homogeneous and isotropic superconductor, $\Delta$ can be set
to be real.

To see clearly the effect of the Fermi surface mismatch on the
Meissner mass, we replace the individual chemical potentials
$\mu_1$ and $\mu_2$ in $\mu=diag(\mu_1,\mu_2)$ by the average
chemical potential $\bar\mu=\left(\mu_1+\mu_2\right)/2$ and the
chemical potential difference
$\delta\mu=\left(\mu_1-\mu_2\right)/2$ and calculate the Meissner
mass of the gauge field as a function of the pairing gap $\Delta$
and the mismatch $\delta\mu$ at fixed $\bar\mu$.

The way to calculate the Meissner mass is quite standard. We
firstly integrate out the fermionic degrees of freedom and then
obtain the effective action for the gauge field. With the
Nambu-Gorkov field defined as $\Psi=(\psi, \psi^\dagger)^T$, the
effective action can be expressed as
\begin{equation}
\label{eff}
S_{eff}[A]=S_A+\frac{\Delta^2}{4G}-\frac{1}{2}\text {
Tr}\ln\left[{\cal S}^{-1}+{\cal A}\right],
\end{equation}
where ${\cal S}^{-1}$ is the inverse of the fermion propagator in
Nambu-Gorkov space
\begin{equation}
{\cal S}^{-1}=\left(\begin{array}{cc}
-\partial_\tau+\frac{\nabla^2}{2m}+\mu&i\Delta\varepsilon
\\
i\Delta\varepsilon&-\partial_\tau-\frac{\nabla^2}{2m}-\mu\end{array}\right),
\end{equation}
and ${\cal A}$ is a matrix related to the gauge field, ${\cal
A}=diag\left({\cal A}^+, {\cal A}^-\right)$ with the elements
${\cal A}^\pm$ defined as
\begin{equation}
{\cal A}^{\pm}=\pm eA^0\mp\frac{e^2}{2m}{\bf
A}^2-\frac{ie}{2m}(\nabla\cdot{\bf A}+{\bf A}\cdot\nabla).
\end{equation}
Since we are interested in the Meissner mass which is related to
the spatial or magnetic component of the gauge field, we need only
the quadratic term in ${\bf A}$ of the effective action,
\begin{equation}
S_{eff}^{(2)}[{\bf A}]=-\frac{1}{2}\sum_k
A^i(-k)\Pi_{ij}(k)A^j(k),
\end{equation}
where $\sum_k=T\sum_n\int{d^3{\bf k}\over (2\pi)^3}$ means fermion
frequency summation and vector momentum integration in the
momentum space $k_\mu=\left(k_0=i\omega_n, {\bf k}\right)$ with
$\omega_n=2n\pi T$ for bosons and $\omega_n=(2n+1)\pi T$ for
fermions. The polarization tensor $\Pi_{ij}$ ($i,j=1,2,3$) can be
explicitly expressed as
\begin{eqnarray}
\Pi_{ij}(k)&=&\frac{e^2}{2}\sum_p\bigg[\frac{\delta_{ij}}{m}\text{Tr}\left({\cal
S}(p)\Sigma\right)\\
&&+\frac{(2p_i-k_i)(2p_j-k_j)}{4m^2}\text{Tr}\left({\cal
S}(p){\cal S}(p-k)\right)\bigg]\nonumber
\label{polar}
\end{eqnarray}
with the matrix $\Sigma$ defined as $\Sigma=diag\left(I_2,
-I_2\right)$, where $I_2$ is the two dimensional identity matrix
in the inner space. The Meissner mass of the gauge field is
defined as
\begin{equation}
m^2_A=\frac{1}{2}\lim_{{\bf k}\rightarrow 0}(\delta_{ij}-k_i
k_j/|{\bf k}|^2)\Pi^{ij}(0,{\bf k}).
\end{equation}
Note that the polarization tensor is composed of two parts: The
first term is the diamagnetic term which is related to the fermion
number density and the second term is the paramagnetic term
including the one-loop diagram of the fermion propagator. This
behavior is quite different from that of a relativistic
system\cite{huang1} where the one-loop diagram contributes to both
diamagnetic and paramagnetic terms. Another convenient way to
calculate the Meissner mass is through the potential
curvature\cite{fukushima,he2},
\begin{equation}
m^2_A=\frac{1}{3}\sum_{i,j=1}^3\delta_{ij}\frac{\partial^2S_{eff}[{\bf
A}]}{\partial A_i\partial A_j}\bigg|_{{\bf A}=0}.
\end{equation}
Obviously, the two methods are equivalent.

Defining the notation
$\epsilon_\Delta=\sqrt{\epsilon_p^2+\Delta^2}$ with
$\epsilon_p={\bf p}^2/(2m)-\bar\mu$, we can explicitly express the
fermion propagator in the Nambu-Gorkov$\otimes$inner space as
\begin{equation}
{\cal S}(p)=\left(\begin{array}{cccccc} {\cal G}_+^+&0&0&-i{\cal
F}_+
\\ 0&{\cal G}_-^+&i{\cal F}_-&0\\ 0&-i{\cal F}_-&{\cal G}_-^-&0 \\ i{\cal
F}_+&0&0&{\cal G}_+^-
\end{array}\right),
\end{equation}
where the nonzero matrix elements are defined as
\begin{eqnarray}
\label{gf}
{\cal
G}_\pm^\mp(p)&=&\frac{i\omega_n\pm\delta\mu\mp\epsilon_p}{(i\omega_n\pm\delta\mu)^2-\epsilon_\Delta^2},\nonumber\\
{\cal F}_\pm(p)
&=&\frac{\Delta}{(i\omega_n\pm\delta\mu)^2-\epsilon_\Delta^2}.
\end{eqnarray}

The Meissner mass squared $m_A^2$ can be decomposed into a
diamagnetic part $\left(m_A^2\right)_d$ and a paramagnetic part
$\left(m_A^2\right)_p$ which come, respectively, from the first
and second terms of the polarization tensor,
\begin{equation}
m_A^2=\left(m_A^2\right)_d+\left(m_A^2\right)_p.
\end{equation}
After a straightforward algebra, they can be evaluated as
\begin{eqnarray}
\left(m_A^2\right)_d &=&\frac{e^2}{2m}\sum_p\Big[\left({\cal
G}_-^++{\cal G}_+^+\right)e^{i\omega_n 0^+}\nonumber\\
&&-\left({\cal G}_-^-+{\cal G}_+^-\right)e^{i\omega_n 0^-}\Big],\nonumber\\
\left(m_A^2\right)_p &=&\frac{e^2}{2m^2}\sum_p\frac{{\bf
p}^2}{3}\Big[{\cal G}_-^+{\cal G}_-^++{\cal G}_+^+{\cal
G}_+^++{\cal G}_+^-{\cal
G}_+^-\nonumber\\
&&+\ {\cal G}_-^-{\cal G}_-^-+2{\cal F}_-{\cal F}_-+2{\cal
F}_+{\cal F}_+\Big].
\end{eqnarray}
Note that we have added the convergent factors in the Matsubara
frequency summation in the diamagnetic part to ensure the
condition $m_A^2(\Delta=0)=0$.

After the Matsubara frequency summation, we obtain
\begin{eqnarray}
\label{t0}
\left(m_A^2\right)_d &=& \frac{e^2}{m}\int_0^\infty
\frac{p^2dp}{2\pi^2}\left[1-{\epsilon_p\over\epsilon_\Delta}\Theta\left(\epsilon_\Delta-\delta\mu\right)\right],\nonumber\\
\left(m_A^2\right)_p &=& -\frac{e^2}{m^2}\int_0^\infty
\frac{p^4dp}{6\pi^2}\delta(\epsilon_\Delta-\delta\mu)
\end{eqnarray}
at $T=0$. The diamagnetic mass is only related to the number
density $n$,
\begin{equation}
\left(m_A^2\right)_d=\frac{ne^2}{m},
\end{equation}
and the paramagnetic mass can be evaluated as
\begin{equation}
\left(m_A^2\right)_p=-\frac{e^2}{m}\frac{\delta\mu\Theta(\delta\mu-\Delta)}
{\sqrt{\delta\mu^2-\Delta^2}}\frac{p_-^3\Theta\left(\mu_-\right)+p_+^3\Theta\left(\mu_+\right)}{6\pi^2}
\end{equation}
with the definitions
\begin{equation}
p_\pm =\sqrt{2m\mu_\pm},\ \ \
\mu_\pm=\bar\mu\pm\sqrt{\delta\mu^2-\Delta^2}.
\end{equation}

In weak coupling region, we have $\bar\mu\gg\delta\mu$ and
\begin{equation}
n\simeq\frac{p_+^3+p_-^3}{6\pi^2}\simeq\frac{(2m\bar\mu)^{3/2}}{3\pi^2},
\end{equation}
and the Meissner mass squared can be approximately expressed as
\begin{equation}
m_A^2=\frac{ne^2}{m}\left[1-\frac{\delta\mu\Theta(\delta\mu-\Delta)}{\sqrt{\delta\mu^2-\Delta^2}}\right].
\end{equation}
In the gapless state with $\delta\mu>\Delta$, the Meissner mass
squared is negative, and there is a discontinuity at the
gapless-gapped transition point $\delta\mu=\Delta$. This
divergence and discontinuity come from the delta function in
(\ref{t0}).

\begin{figure}[!htb]
\begin{center}
\includegraphics[width=8cm]{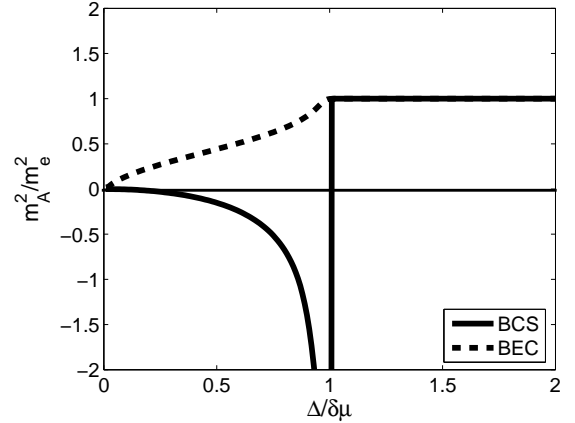}
\caption{The Meissner mass squared, scaled by $m_e^2=ne^2/m$, as a
function of $\Delta/\delta\mu$. We take $\bar\mu=0.99\epsilon_F,
\Delta=0.05\epsilon_F$ in the BCS case and
$\bar\mu=-0.1\epsilon_F, \Delta=0.5\epsilon_F$ in the BEC case
with $\epsilon_F$ being the Fermi energy for free fermions.}
\label{fig1}
\end{center}
\end{figure}

When the coupling becomes strong enough, the system will enter the
BEC region and we have $\bar\mu<0$. In this case there is only one
gapless node $p_+$ and the Meissner mass squared reads
\begin{equation}
m_A^2=\frac{ne^2}{m}\left[1-\frac{\delta\mu\Theta(\delta\mu-\Delta)}
{\sqrt{\delta\mu^2-\Delta^2}}\frac{p_+^3\Theta\left(\mu_+\right)}{6\pi^2n}\right].
\end{equation}
For $\bar{\mu}<0$, $p_+$ becomes much smaller than the Fermi
momentum $p_F=(3\pi^2n)^{1/3}$, the Meissner mass squared in the
gapless phase can be positive\cite{pao,gubankova}, and the
divergence at the gapless-gapped transition point disappears due
to the fact that at $\delta\mu=\Delta$ both $p_+$ and $p_-$ become
imaginary. In Fig.\ref{fig1}, we plot the Meissner mass squared in
BCS and BEC cases. In BCS case, the gapless state suffers magnetic
instability, while in the strong coupling BEC region, the
instability is fully cured.

\section {Non-Abelian Symmetry}
\label{s4}
In this section we consider a system with two types of inner
degrees of freedom which we call color and flavor. They can also
be interpreted as spin, isospin and hyperfine state in different
systems. Suppose there are two flavors and three colors, like two
flavor QCD, the system can be modelled by the partition function
\begin{eqnarray}
Z&=&\int[d\psi^\dagger][d\psi][dA]e^{S_A}\times\\
&&e^{\int_x\left[\psi^\dagger\left(-D_\tau+\frac{{\bf
D}^2}{2m}+\mu\right)
\psi+G(\psi^\dagger\varepsilon\epsilon^{c}\psi^*)(\psi^T\varepsilon\epsilon^{c}\psi)\right]},\nonumber
\end{eqnarray}
where $(\varepsilon)_{\alpha\beta}\equiv\varepsilon_{\alpha\beta}$
and $(\epsilon^c)^{ab}\equiv\epsilon^{abc}$ are antisymmetric
matrices in the flavor and color spaces with the flavor index
denoted by $\alpha,\beta$ and color by $a,b,c$. We have introduced
a SU(3) gauge field $A_\mu^a$ corresponding to the color degree of
freedom via the derivative $D_\mu$ in (\ref{gauge}).

We consider pairing only between fermions with different flavors
and colors and introduce the order parameter
\begin{equation}
\Delta = -2G\langle\psi^T\varepsilon\epsilon^{3}\psi\rangle
\end{equation}
which spontaneously breaks the symmetry from SU(3) to SU(2), where
we have assumed that only the first two colors participate in the
condensate, while the third one does not. Similar to the treatment
in Abelian case, we obtain the effective action (\ref{eff}) for
the gauge field with the inverse fermion propagator in
Nambu-Gorkov space
\begin{equation}
{\cal S}^{-1}=\left(\begin{array}{cc}
-\partial_\tau+\frac{\nabla^2}{2m}+\mu&\Delta\varepsilon\epsilon^{3}
\\ \Delta\varepsilon\epsilon^{3}&-\partial_\tau-\frac{\nabla^2}{2m}-\mu\end{array}\right)
\end{equation}
and the gauge field elements
\begin{equation}
{\cal A}^{\pm}=\pm gT_aA_a^0\mp\frac{g^2T_a^2}{2m}{\bf
A}_a^2-\frac{igT_a}{2m}(\nabla\cdot{\bf A}_a+{\bf
A}_a\cdot\nabla).
\end{equation}

If the fermion chemical potential is color independent, we have
$\mu=diag\left(\bar\mu+\delta\mu, \bar\mu-\delta\mu\right)\otimes
I_3$ with $I_3$ being the three dimensional identity matrix in the
color space, the Fermi surface mismatch between the pairing
fermions does not break the $SU(3)$ gauge symmetry explicitly, and
the quadratic term of the effective action for the magnetic
component of the gauge field reads
\begin{equation}
S_{eff}^{(2)}[{\bf A}]=-\frac{1}{2}\sum_k
A^i_a(-k)\Pi_{ab}^{ij}(k)A^j_b(k)
\end{equation}
with the polarization tensor
\begin{eqnarray}
\Pi_{ab}^{ij}(k)&=&\frac{g^2}{2}\sum_p\bigg[\frac{\delta_{ij}\delta_{ab}}{m}
\text {Tr}[{\cal
S}(p)\Sigma_a]\\
&&+\frac{(2p_i-k_i)(2p_j-k_j)}{4m^2}\text{Tr}[{\cal
S}(p)\Gamma_a{\cal S}(p-k)\Gamma_b]\bigg],\nonumber
\end{eqnarray}
where the matrices $\Sigma_a$ and $\Gamma_a$ are defined as $
\Sigma_a = diag\left(I_2\otimes T_a^2, -I_2\otimes T_a^2\right)$
and $\Gamma_a =diag\left(I_2\otimes T_a, I_2\otimes T_a\right)$.

In the 12-dimensional Nambu-Gorkov$\otimes$flavor$\otimes$color
space, the fermion propagator can be explicitly expressed as
\begin{widetext}
\begin{eqnarray}
{\cal S}(p)&=&\left(\begin{array}{cccccccccccc} {\cal
G}_+^+&0&0&0&0&0&0&0&0&0&-{\cal F}_+&0
\\ 0&{\cal G}_+^+&0&0&0&0&0&0&0&{\cal F}_+&0&0\\ 0&0&{\cal G}_{0+}^-&0&0&0&0&0&0&0&0&0
\\ 0&0&0&{\cal G}_-^+&0&0&0&{\cal F}_-&0&0&0&0 \\ 0&0&0&0&{\cal G}_-^+&0&-{\cal F}_-&0&0&0&0&0 \\
0&0&0&0&0&{\cal G}_{0-}^-&0&0&0&0&0&0\\0&0&0&0&-{\cal F}_-&0&{\cal
G}_-^-&0&0&0&0&0\\0&0&0&{\cal F}_-&0&0&0&{\cal
G}_-^-&0&0&0&0\\0&0&0&0&0&0&0&0&{\cal G}_{0-}^+&0&0&0\\0&{\cal
F}_+&0&0&0&0&0&0&0&{\cal G}_+^-&0&0\\-{\cal
F}_+&0&0&0&0&0&0&0&0&0&{\cal G}_+^-&0\\0&0&0&0&0&0&0&0&0&0&0&{\cal
G}_{0+}^+
\end{array}\right)
\end{eqnarray}
\end{widetext}
with the elements ${\cal G}^\mp_\pm, {\cal F}_\pm$ shown in
(\ref{gf}) and the propagator ${\cal G}^\mp_{0\pm}$ for the
unpaired fermions defined as
\begin{equation}
{\cal G}_{0\pm}^\mp=\frac{1}{i\omega_n\pm\delta\mu\mp\epsilon_p}.
\end{equation}
Since in the phase with nonzero condensate there is still the
symmetry SU(2), only the generators $T_4,T_5,T_6,T_7,T_8$ are
broken, we need to calculate the Meissner mass squared only for
$a=4,5,6,7,8$.

With the definition for the Meissner mass squared in non-Abelian
case,
\begin{equation}
\left(m^2_A\right)_{ab}=\frac{1}{3}\sum_{i,j=1}^3\delta_{ij}\frac{\partial^2S_{eff}[{\bf
A}]}{\partial A_i^a\partial A_j^b}\bigg|_{{\bf A}=0},
\end{equation}
we first consider $a=4,5,6,7$. It is straightforward to see that
all the off-diagonal elements of the matrix
$\left(m_A^2\right)_{ab}$ in the color subspace vanish and the
diagonal elements are the same,
\begin{equation}
\left(m^2_A\right)_{44}=\left(m_A^2\right)_{55}=\left(m_A^2\right)_{66}=\left(m_A^2\right)_{77}.
\end{equation}
Therefore, we need to evaluate $m_{A,4}^2\equiv
\left(m_A^2\right)_{44}$ only. Taking into account the
decomposition,
\begin{equation}
m_{A,4}^2=\left(m_{A,4}^2\right)_d+\left(m_{A,4}^2\right)_p,
\end{equation}
the diamagnetic and paramagnetic parts coming, respectively, from
the first and second terms of the polarization tensor read
\begin{eqnarray}
\left(m_{A,4}^2\right)_d &=&\frac{g^2}{8m}\sum_p\Big[\left({\cal
G}_{0-}^-+{\cal G}_{0+}^-+{\cal G}_-^++{\cal
G}_+^+\right)e^{i\omega_n0^+}\nonumber\\
&&-\left({\cal G}_{0-}^++{\cal G}_{0+}^++{\cal G}_-^-+{\cal G}_+^-\right)e^{i\omega_n0^-}\Big],\nonumber\\
\left(m_{A,4}^2\right)_p &=& \frac{g^2}{4m^2}\sum_p\frac{{\bf
p}^2}{3}\Big[{\cal G}_{0-}^+{\cal
G}_-^-+{\cal G}_{0-}^-{\cal G}_-^+\nonumber\\
&& +\ {\cal G}_{0+}^+{\cal G}_+^-+{\cal G}_{0+}^-{\cal
G}_+^+\Big].
\end{eqnarray}
Note that only the loops with an unpaired fermion and a paired
fermion contribute to the paramagnetic term. After the Matsubara
frequency summations, we obtain
\begin{widetext}
\begin{eqnarray}
\left(m_{A,4}^2\right)_d &=& \frac{g^2}{2m}\int_0^\infty
\frac{p^2dp}{2\pi^2}\left[\Theta(\delta\mu-\epsilon_p)+
\Theta(-\delta\mu-\epsilon_p)+\frac{\epsilon_p}{\epsilon_\Delta}\Theta(\delta\mu-\epsilon_\Delta)+
\left(1-\frac{\epsilon_p}{\epsilon_\Delta}\right)\right],\nonumber\\
\left(m_{A,4}^2\right)_p &=& -\frac{g^2}{m^2}\int_0^\infty
\frac{p^4dp}{3\pi^2}\left[\frac{\epsilon_p}{\Delta^2}\left(\Theta(\delta\mu-\epsilon_p)+\Theta(-\delta\mu-\epsilon_p)\right)
+\left(\frac{1}{\epsilon_\Delta+\epsilon_p}-\frac{1}{2\epsilon_\Delta}\right)
-\left(\frac{\epsilon_\Delta}{\Delta^2}-\frac{1}{2\epsilon_\Delta}\right)\Theta(\delta\mu-\epsilon_\Delta)\right]\nonumber
\end{eqnarray}
\end{widetext}
at $T=0$. At weak coupling, the integrals can be approximately
evaluated and the final result is simplified as
\begin{equation}
m_{A,4}^2=2m_g^2\left[\frac{1}{2}-\frac{\delta\mu^2}{\Delta^2}+\frac{\delta\mu^2}{\Delta^2}
\frac{\sqrt{\delta\mu^2-\Delta^2}}{\delta\mu}\Theta(\delta\mu-\Delta)\right],
\end{equation}
where $m_g^2=ng^2/m$ is a $\Delta$ and $\delta\mu$ independent
constant.

\begin{figure}[!htb]
\begin{center}
\includegraphics[width=8cm]{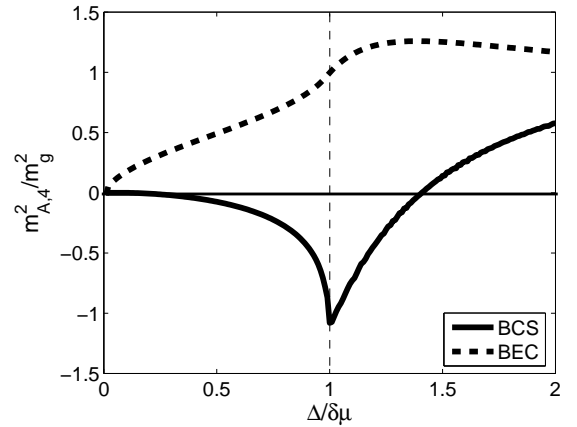}
\caption{The Meissner mass squared $m_{A,4}^2$, scaled by
$m_g^2=ng^2/m$, as a function of $\Delta/\delta\mu$. We take
$\bar\mu=0.99\epsilon_F, \Delta=0.05\epsilon_F$ in the BCS case
and $\bar\mu=-0.1\epsilon_F, \Delta=0.5\epsilon_F$ in the BEC
case. } \label{fig2}
\end{center}
\end{figure}

In Fig.\ref{fig2}, we show the Meissner mass squared $m_{A,4}^2$
as a function of the gap parameter $\Delta$. In the BCS case, both
the gapless state in the region $\Delta/\delta\mu<1$ and the
gapped state in the region $1<\Delta/\delta\mu<\sqrt{2}$ suffer
magnetic instability, and the Meissner mass squared is continuous
at $\Delta/\delta\mu=1$. The continuity at the gapless-gapped
transition point is totally different from the case with Abelian
symmetry. We also observed that the magnetic instability can be
cured in the strong coupling BEC region.

Now we turn to the calculation of the Meissner mass squared with
$a=8$. Again we make the diamagnetic and paramagnetic
decomposition,
\begin{equation}
m_{A,8}^2\equiv
\left(m_A^2\right)_{88}=\left(m_{A,8}^2\right)_d+\left(m_{A,8}^2\right)_p,
\end{equation}
and each part can further be divided into a unpaired fermion term
and a paired fermion term,
\begin{eqnarray}
\left(m_{A,8}^2\right)_d &=&
\left(m_{A,8}^2\right)_{d}^0+\left(m_{A,8}^2\right)_{d}^\Delta,\nonumber\\
\left(m_{A,8}^2\right)_p &=&
\left(m_{A,8}^2\right)_{p}^0+\left(m_{A,8}^2\right)_{p}^\Delta
\end{eqnarray}
with the unpaired fermion contributions
\begin{eqnarray}
\left(m_{A,8}^2\right)_d^0 &=&
\frac{g^2}{8m}\sum_p\frac{4}{3}\Big[\left({\cal G}_{0-}^-+{\cal
G}_{0+}^-\right)e^{i\omega_n0^+}\nonumber\\
&&-\left({\cal G}_{0-}^++{\cal
G}_{0+}^+\right)e^{i\omega_n0^-}\Big],\nonumber\\
\left(m_{A,8}^2\right)_p^0 &=& \frac{g^2}{8m^2}\sum_p\frac{{\bf
p}^2}{3}\frac{4}{3}\Big[{\cal G}_{0-}^-{\cal G}_{0-}^-+{\cal
G}_{0+}^-{\cal
G}_{0+}^-\nonumber\\
&&+{\cal G}_{0-}^+{\cal G}_{0-}^++{\cal G}_{0+}^+{\cal
G}_{0+}^+\Big],
\end{eqnarray}
and the paired fermion contributions
\begin{eqnarray}
\left(m_{A,8}^2\right)_d^\Delta &=&
\frac{g^2}{8m}\sum_p\frac{2}{3}\Big[\left({\cal G}_-^++{\cal
G}_+^+\right)e^{i\omega_n0^+}\nonumber\\
&&-\left({\cal G}_-^-+{\cal
G}_+^-\right)e^{i\omega_n0^-}\Big],\nonumber\\
\left(m_{A,8}^2\right)_p^\Delta &=&
\frac{g^2}{8m^2}\sum_p\frac{{\bf p}^2}{3}\frac{2}{3}\Big[{\cal
G}_-^+{\cal G}_-^++{\cal G}_+^+{\cal G}_+^++{\cal G}_+^-{\cal
G}_+^-\nonumber\\
&&+{\cal G}_-^-{\cal G}_-^-+2{\cal F}_-{\cal F}_-+2{\cal F}_+{\cal
F}_+\Big].
\end{eqnarray}
The terms $\left(m_{A,8}^2\right)_{d}^0$ and
$\left(m_{A,8}^2\right)_{p}^0$ cancel to each other,
\begin{equation}
\left(m_{A,8}^2\right)_{d}^0+\left(m_{A,8}^2\right)_{p}^0=0,
\end{equation}
which means that the unpaired fermions have no contribution to the
Meissner effect corresponding to the 8th gauge field, and the
final expression of $m_{A,8}^2$ is the same as that in the Abelian
case except for a constant factor,
\begin{equation}
m_{A,8}^2=\frac{g^2}{6e^2}m_A^2.
\end{equation}
Therefore, the magnetic stability analysis for the 8th gluon
should be the same as the one in the Abelian model.

\section {Summary}
\label{s5}
We have investigated the origin of two types of magnetic
instabilities induced by the Fermi surface mismatch between the
two pairing fermions in a U(1) Abelian model and a SU(3)
non-Abelian model. Our results show that the two types of
instabilities are very different. The Abelian instability occurs
only in the gapless state where the chemical potential mismatch
$\delta\mu$ is larger than the gap $\Delta$, and the Meissner mass
squared becomes divergent at the gapless-gapped transition point
$\delta\mu=\Delta$. However, the non-Abelian instability occurs in
both gapless and gapped states, and there is no singularity at the
gapless-gapped transition point. In non-Abelian systems, there are
both Abelian and non-Abelian magnetic instabilities. The former
corresponds to the broken diagonal generators and the latter to
those broken off-diagonal generators. While only the paired
fermions contribute to the Abelian instability, both paired and
unpaired fermions contribute to the non-Abelian instability which
leads to the continuity of the Meissner mass squared at the
gapless-gapped transition point.

Since the non-relativistic model we considered in this paper has
the simple but essential pairing structure and good ultraviolet
behavior, it is quite convenient to study the non-Abelian LOFF
state\cite{fukus,hashimoto} and discuss the competition between
the Abelian and non-Abelian LOFF states\cite{kiriyama}.

Our investigation indicates that the magnetic instability induced
by mismatched fermi surfaces does not depend on the details of the
attractive interaction and whether the system is relativistic or
non-relativistic. The instability is controlled by the symmetry of
the system. While a relativistic and a non-relativistic Fermi gas
may have different dynamics, their magnetic instabilities are very
similar if they have the same symmetry. For instance, the
chromomagnetic instability discovered in relativistic color
superconductor may be closely related to the experiments in
non-relativistic and non-Abelian condensed matter such as the
three-component Fermi gas, and other new discoveries in the study
of color superconductivity, such as the abnormal number of
Nambu-Goldstone bosons\cite{blaschke,he3,ebert} and the
instability induced by the mismatch between paired and unpaired
fermions\cite{he4}, may be realized in condensed matter systems.

{\bf Acknowledgments:}\  The work was supported in part by the
grants NSFC10575058, 10425810, 10435080 and SRFDP20040003103.

\end{document}